\documentclass[twocolumn,aps,prl,amsmath,amssymb]{revtex4}

\usepackage{graphicx}

\usepackage{dcolumn}

\usepackage{bm}
\usepackage{amsmath,tabularx} 

\usepackage{epstopdf}

\DeclareGraphicsExtensions{.pdf,.eps,.png,.jpg,.mps}

\begin{document}

\preprint{AIP/123-QED}

\title{Graphene-Silicon-On-Insulator (GSOI) Schottky Diode Photodetectors}

\author{H. Selvi$^{1}$}

\author{E.W. Hill$^{2,4}$}

\author{P. Parkinson$^{3,5}$}

\author{T.J. Echtermeyer$^{1,2,3}$}

\email{tim.echtermeyer@manchester.ac.uk}

\affiliation{$^1$School of Electrical \& Electronic Engineering, University of Manchester, Manchester M13 9PL, UK}

\affiliation{$^2$National Graphene Institute, University of Manchester, Manchester M13 9PL, UK}

\affiliation{$^3$Photon Science Institute, University of Manchester, Manchester M13 9PL, UK}

\affiliation{$^4$Manchester Centre For Mesoscience and Nanotechnology, University of Manchester, Manchester M13 9PL, UK}

\affiliation{$^5$School of Physics and Astronomy, University of Manchester, Manchester M13 9PL, UK}

\begin{abstract}

Graphene-silicon (GS) Schottky junctions have been demonstrated as an efficient architecture for photodetection. However, the response speed of such devices for free space light detection has so far been limited to 10's-100's of kHz for wavelength $\lambda >$ 500nm. Here, we demonstrate graphene-silicon Schottky junction photodetectors fabricated on a silicon-on-insulator substrate (SOI) with response speeds approaching 1GHz, attributed to the reduction of the photo-active silicon layer thickness to 10$\mu$m and with it a suppression of speed-limiting diffusion currents. Graphene-silicon-on-insulator photodetectors (GSOI-PDs) exhibit a negligible influence of wavelength on response speed and only a modest compromise in responsivities compared to GS junctions fabricated on bulk silicon. Noise-equivalent-power (NEP) and specific detectivity (D$^*$) of GSOI photodetectors are 14.5pW and 7.83$\times10^{\rm{10}}$ cm Hz$^{\rm{1/2}}$W$^{\rm{-1}}$, respectively, in ambient conditions. We further demonstrate that combining GSOI-PDs with micro-optical elements formed by modifying the surface topography enables engineering of the spectral and angular response.

\end{abstract}

\maketitle
Graphene is an appealing material to realize ultrafast and broadband photodetectors (PDs) due to its versatile electronic and optical properties such as high carrier mobility\cite{mayorov2011interaction}, ultrafast carrier dynamics\cite{brida2013ultrafast} and broadband optical absorption\cite{kuzmenko2008universal,nair2008fine}. However, graphene's low optical absorption of $\sim$2.3\% and small photo-active area of devices are a bottleneck for photodetectors solely based on graphene. Photoresponsivities of graphene-based photodetectors based on metal-graphene-metal architectures (MGM) are typically below 10mAW$^{\rm{-1}}$ for visible (VIS) and near-infrared (NIR) wavelengths\cite{mueller2010graphene,xia2009ultrafast,lee2008contact} and the photoactive area is restricted to the periphery ($\sim$100-200nm) of the graphene contact\cite{mueller2009role,echtermeyer2016surface}. Various strategies have been exploited to enhance light absorption in graphene, e.g. by integrating graphene into optical microcavities\cite{engel2012light,furchi2012microcavity} or combining graphene with plasmonic nanostructures\cite{echtermeyer2011strong,fang2012graphene}. More recently, graphene-silicon (GS) Schottky junction photodiodes have been demonstrated as an efficient platform for photodetection\cite{amirmazlaghani2013graphene,wang2013high,an2013tunable,lv2013high,an2013metal,liu2014quantum,chen2015high,goykhman2016chip,li2016high,srisonphan2016hybrid,di2016tunable,riazimehr2016spectral,wan2017self,shen2017high,riazimehr2017high,di2017hybrid,tao2017hybrid,selvi2018towards,casalino2017vertical} and photovoltaic\cite{li2015carbon,li2010graphene,miao2012high,an2013optimizing,lin2013graphene,song2015role} applications. The co-integration of graphene with silicon technology allows the realization of a hybrid platform that is suitable for large scale fabrication due to the possible integration of graphene into back end-of-line (BEOL) complementary metal-oxide-semiconductor (CMOS) processing\cite{pospischil2013cmos,pasternak2016graphene}. 

GS Schottky PDs exhibit a dual operating regime where both graphene and silicon act as active light absorbing materials for different wavelength ranges\cite{riazimehr2016spectral}. Devices can show high responsivities on the order of hundreds of mA/W, comparable to commercial silicon photodiodes for wavelength ranges with photon energies above the silicon bandgap ($\lambda$ \textless \ 1.1$\mu$m) which is facilitated by the high optical transmittance of graphene ($\sim$97.3$\%$). Detection of light with energies below the silicon band gap is enabled by the broadband absorption of graphene\cite{kuzmenko2008universal} and responsivities reduce to a few mAW$^{\rm{-1}}$ or less for $\lambda >$ 1.1$\mu$m \cite{selvi2018towards}. One important advantage of GS devices is their large photoactive area due to the vertical nature of the graphene-silicon junction as opposed to the lateral junction as in MGM photodetectors.

To date, the performance of GS PDs in terms of response speed remains far below that of MGM and commercial silicon photodetectors which exhibit cut-off frequencies (f$_c$) on the order of GHz for free space light detection\cite{urich2011intrinsic,sze2007physics}. Conversely, GS PDs demonstrated response speeds of 10's-100's of kHz for wavelengths $\lambda >$ 500nm \cite{an2013tunable,lv2013high,li2016high,chen2015high,shen2017high,tao2017hybrid,selvi2018towards}. This limitation can be attributed to speed-limiting diffusion currents due to photo-generated carriers deep within the silicon substrate of typically employed thicknesses ($\sim$ 500$\mu$m). A solution to improve the response speed of GS devices is replacing commonly employed bulk silicon with a silicon-on-insulator (SOI) substrate. SOI, a thin silicon layer on top of a buried silicon oxide layer (BOX), is frequently used to implement CMOS electronics because of many advantages compared to their bulk silicon counterparts; SOI allows full dielectric isolation between neighboring devices, reduced leakage currents and reduced capacitive coupling, full depletion of the active silicon layer and 3-dimensional device architectures\cite{colinge2004soi} which makes SOI suitable for high speed and low power applications. SOI substrates additionally provide processing advantages since the BOX acts as a well-defined etch-stop and substrates can further be employed for micro-electro-mechanical-systems (MEMS)\cite{dao2010mems}.

\begin{figure*}[htbp]
\centering{
\includegraphics[width=150mm]{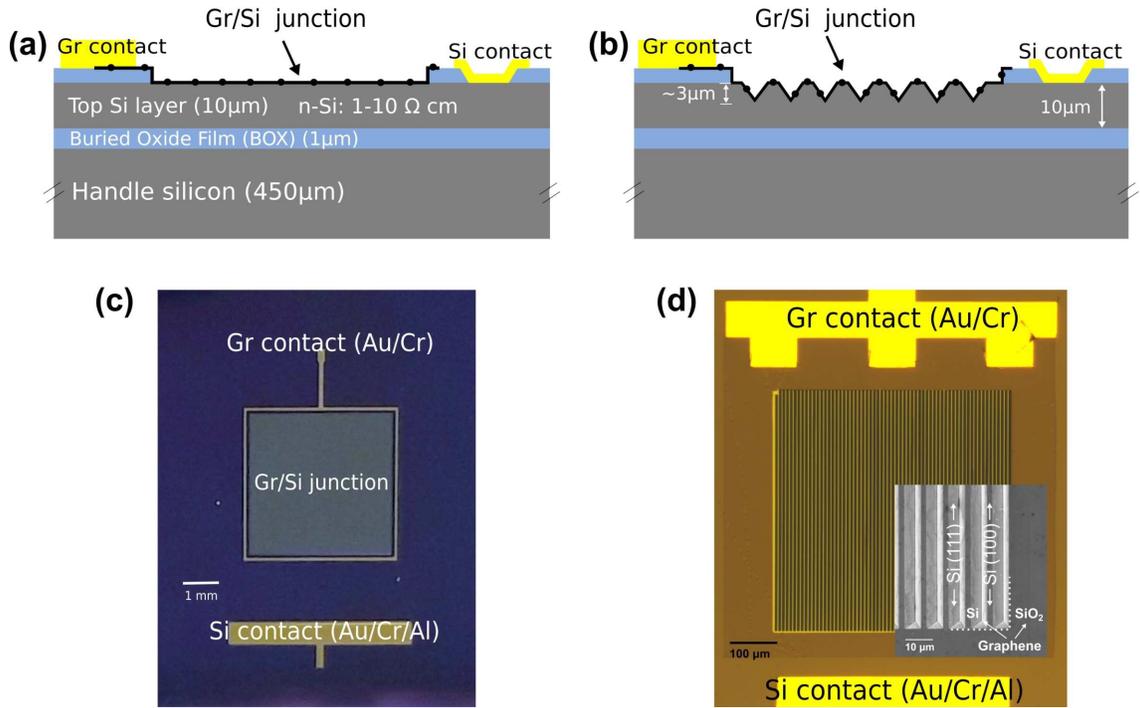}}
\caption{a) Schematic structure of the GSOI-planar device where graphene is transferred on the planar silicon (100) surface. b) Schematic structure of the GSOI-grating device where graphene is transferred on the structured silicon surface that consist of an array of V-grooves with (111) silicon facets separated by unpatterned planar (100) silicon lines of 3$\mu$m width. The V-grooves are 3$\mu$m deep and 5.5$\mu$m wide. c,d) Optical images of the fabricated devices. The active area of the GSOI-planar device is 4$\times$4mm$^2$ and for the GSOI-grating is 0.5$\times$0.5mm$^2$. Inset: Scanning electron micrograph (SEM) image of the grating structure formed by the V-grooves.}
\label{fig:process}
\end{figure*}

SOI wafers with a 10$\mu$m thick, lightly-doped (1-10 $\Omega$cm) n-type silicon device layer (100) were employed to fabricate two sets of devices, a planar GSOI Schottky diode (GSOI-planar) and GSOI Schottky diode with a grating surface topography (GSOI-grating). The architecture of both device types is shown in Fig.\ref{fig:process}. For the GSOI-planar device, graphene is transferred onto the flat (100) oriented silicon surface. For the GSOI-grating device, graphene is transferred onto an array of self-terminated V-grooves formed by anisotropic wet etching of silicon in KOH solution, exposing both the (100)and (111) silicon surfaces with an angle of 54.7$^\circ$ between them\cite{petersen1982si}. To pattern the silicon surface of the GSOI-grating device, 100nm of SiN$_{\rm{x}}$ were deposited on the silicon surface by inductively coupled plasma enhanced chemical vapor deposition (ICP-PECVD, Oxford Instruments). A grating structure consisting of an array of trenches ($\sim$5.5$\mu$m wide) is patterned in the SiN$_{\rm{x}}$ layer employing optical lithography and subsequent dry etching (reactive ion etching). This grating structure formed in the SiN$_{\rm{x}}$ layer then serves as a hard mask for the anisotropic etching of silicon along the grating trenches carried out in 30wt$\%$ KOH solution at 80$^{\circ}$C. As a result, 3$\mu$m deep (111)-self-terminated V-grooves are etched into the silicon surface. The substrate is then dipped in acetic acid to remove the KOH residues and the SiN$_{\rm{x}}$ mask is completely removed by dry etching.

After the substrate fabrication for the GSOI-grating device, the device fabrication process is identical for both the GSOI-planar and GSOI-grating devices. The previously fabricated grating substrate and un-processed substrate (planar) are both coated with a 100nm thick silicon-dioxide (SiO$_{\rm{2}}$) layer by ICP-PECVD (Oxford Instruments). Subsequently, two areas are opened in the SiO$_{\rm{2}}$ layer for the formation of the GSOI junction and the metal-silicon contact. These areas are defined by optical lithography and the oxide layer in these areas is removed by wet etching in a buffer-oxide-etch (BOE) solution. In the case of the GSOI-grating device, the GSOI junction area corresponds to the previously patterned region. Right after the BOE of the oxide, contacts to the top silicon layer are fabricate by an additional lithography step followed by thermal evaporation of Au/Cr/Al (50nm/3nm/50nm) and lift-off. Aluminum is chosen as the metal contact to the silicon substrate as it forms an ohmic contact to silicon due to its low work function\cite{card1976al}. The additional layers of Au/Cr on top of Al serve as protective layers for Al during subsequent BOE etching processes.

Commercially sourced graphene grown by chemical vapour deposition (CVD) on copper foil is spin coated with a 200nm thick poly-methyl-methacrylate (PMMA, 950K A4) layer before being etched in ammonium persulfate solution to remove the copper foil. The PMMA/graphene membrane is rinsed in de-ionised (DI) water twice before the final transfer to remove etchant and copper residues. The substrates are dipped in BOE for five seconds to remove any native oxide layer formed on the silicon area for the graphene-silicon junction window, before the wet transfer of CVD graphene. The substrates with the PMMA/graphene membrane are baked at 140$^{\circ}$C for 5 minutes to enhance graphene adhesion to the substrate before removing the PMMA layer in an acetone bath. The shape of graphene is defined by optical lithography and subsequent O$_{\rm{2}}$/Ar plasma etching (60s at 8W, Moorfield NanoETCH). A further optical lithography step is used to define the electrical contact to graphene and Au/Cr (50nm/3nm) are evaporated as contact metals with a following lift off step. Optical images of the finalized devices are shown in Fig. \ref{fig:process}c,d). The quality of graphene on top of SiO$_{\rm{2}}$ and on top of the silicon window is examined by Raman spectroscopy under 532nm laser excitation after completion of the device fabrication. Obtained Raman spectra (ESI) show a negligible D peak and a single Lorentzian 2D peak with a full width at half maximum (FWHM) below 35cm$^{-1}$, confirming the successful transfer of monolayer CVD-grown graphene.

\begin{figure}[htbp]
\centering{
\includegraphics[width=80mm]{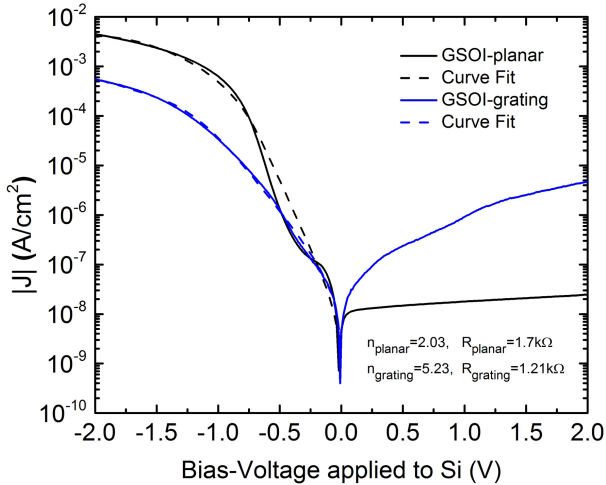}}
\caption{J-V characteristics of the GSOI-planar and GSOI-grating devices in the dark and the fitted Shockley equation (eq.\ref{eq:Shockley}).}
\label{fig:JV}
\end{figure}

The current density-voltage (J-V) characteristics of both the GSOI-planar and GSOI-grating devices are shown in Fig.\ref{fig:JV}. Devices were tested under dark conditions, at room temperature and in ambient atmosphere. Both device types exhibit a rectifying behavior with current on- to off-ratios (I$_{\rm{on}}$/I$_{\rm{off}}$, at $\pm2V$), exceeding 10$^5$ and 10$^2$ for the GSOI-planar and  the GSOI-grating device, respectively. It is noteable that the GSOI-grating device exhibits an increased dark current density under reverse bias compared to the GSOI-planar device. The exact reason for this is unknown so far but we suspect that both the inhomogenous doping profile of graphene due to the patterned substrate as well as the differences in silicon surfaces in contact with graphene, (100) vs (111), will play a role. The J-V characteristic of an ideal diode can be described by the Shockley equation\cite{sze2007physics} which allows the extraction of the ideality factor (n) and series resistance (R$_{\rm{s}}$)

\begin{equation}
\rm{I} = \rm{A}\rm{A}^*\rm{T}^2\rm{exp}\left(\frac{-q\phi_{\rm{B}}}{k_{\rm{B}}\rm{T}}\right)\left[\rm{exp}(\frac{q(V-\rm{IR}_{\rm{S}})}{nk_{\rm{B}}\rm{T}})-1\right]
\label{eq:Shockley}
\end{equation}

Here, $A$ is the Schottky diode contact area, $A^*$ is the effective Richardson constant (112 A$^{-2}$ K$^{-2}$) for n-type silicon) and $\phi_{\rm{B}}$ is the Schottky barrier height (SBH) for a given voltage (V), $k_{\rm{B}}$ is the Boltzmann constant, $q$ the electron charge, and $T$ is the temperature in Kelvin. The values of n and R$_{\rm{s}}$ obtained through fitting eq.\ref{eq:Shockley} to the experimental data are 2.03 and 1.7k$\Omega$ for the SOI-planar device, respectively, and 5.23 and 1.21k$\Omega$ for the SOI-grating device, respectively. We intentionally employ a fit of eq.\ref{eq:Shockley} over the whole bias voltage range despite deviations of the experimental data from ideal diode characteristics. This avoids vastly different results for n and R$_{\rm{s}}$ depending on which voltage region the experimental data has been fitted to. It further allows a fair comparison with other devices since it prevents choosing a sweet spot in the J-V characteristics to optimize n and R$_{\rm{s}}$. The extraction method is described in detail in\cite{cheung1986extraction}.

\begin{figure}[htbp]
\centering{
\includegraphics[width=85mm]{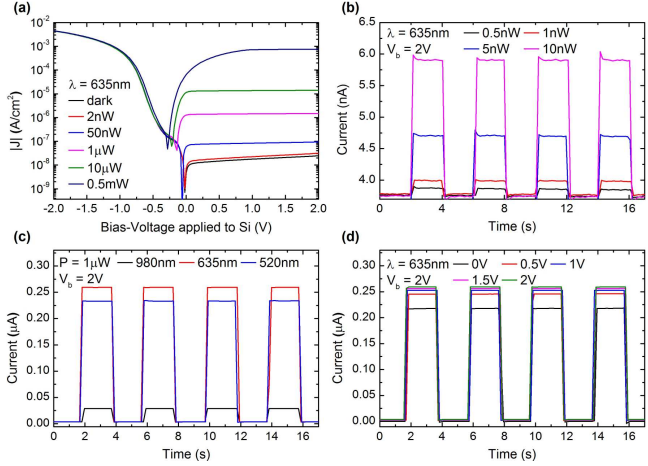}}
\caption{ Photoresponse of the GSOI-planar device in various conditions. a) J-V characteristics under different light powers (P = 2nW..0.5mW) at $\lambda$ = 635nm wavelength. b) Transient measurement under decreasing light powers (P = 10nW..0.5nW) at $\lambda$ = 635nm wavelength at V$_b$ = 2V reverse bias. c) Photocurrent response of the device for different wavelengths at constant light power (P = 1$\mu$W) under V$_b$ = 2V reverse bias. d) Photocurrent response of the device at different reverse biases (V$_b$ = 0..2V) at $\lambda$ = 635nm intermittent illumination (P = 1$\mu$W).}
\label{fig:light}
\end{figure}

Subsequently, J-V measurements of both device types have been carried out under optical excitation at different wavelengths and varying optical powers. Fig.\ref{fig:light}a) shows the J-V characteristics of the GSOI-planar device under illumination with $\lambda$ = 635nm laser light at optical powers varying from P = 2nW..0.5mW (device area 4$\times$4mm$^2$) . As expected for a well-behaved photodiode, the forward current remains unchanged under illumination, however, the reverse current increases with increasing incident light powers due to generated photocurrents at the junction \cite{selvi2018towards}. The time dependent measurement of the current of the GSOI-planar device under intermittent illumination at a reverse bias of V$_b$ = 2V (Fig.\ref{fig:light}b)) shows that light powers as low as 500pW, corresponding to an intensity of $\sim$ 3nW/cm$^2$, can be clearly detected. It demonstrates the high sensitivity of the GSOI-planar device due to the low dark current of $\sim$ 4nA ($\sim$ 25nA/cm$^2$). Fig.\ref{fig:light}c) shows transient measurements conducted with different laser wavelengths at constant power (P = 1$\mu$W) when the device is operated at V$_b$ = 2V reverse bias. The responsivity of the device is 0.23A/W, 0.26A/W and 0.029A/W for 520, 635 and 980nm laser light, respectively. Fig.\ref{fig:light}d) demonstrates the bias voltage dependence of the photoresponse of the GSOI-planar device under $\lambda$ = 635nm laser light illumination. Higher reverse biases increase the depletion width of the Schottky junction and increase the electric field in the depletion region which facilitates separation of photogenerated carries\cite{selvi2018towards,sze2007physics}.

Sensitivity is an important metric that represents the performance of a photodetector by quantifying the ultimate capability to detect weakest signals. Achieving high sensitivity requires both large responsivity R and low noise. Noise-equivalent-power (NEP) and specific detectivity (D$^*$) quantify the detection limit of a photodetector and are defined as\cite{gong2009detectivity}

\begin{figure}[htbp]
\centering{
\includegraphics[width=85mm]{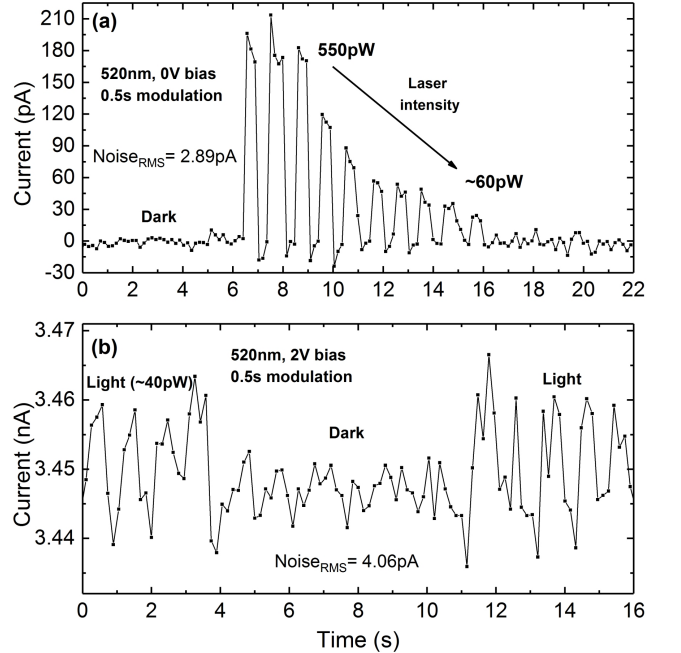}}
\caption{Time dependent current of the GSOI-planar device under low light power in ambient condition. The device is intermittently illuminated with laser light of $\lambda$ = 520nm wavelength of sub-nanowatt intensity to characterize Noise$_{\rm{RMS}}$, NEP and D* under different reverse bias voltages of a) V$_b$ = 0V and b) V$_b$ = 2V.}
\label{fig:noise}
\end{figure}

    \noindent\begin{minipage}{40mm}
\begin{equation}
\rm{NEP}= \frac{\rm{Noise}_{\rm{RMS}}}{\rm{R}} 
\end{equation}
    \end{minipage}%
\begin{minipage}{40mm}
\begin{equation}
\rm{D}^* = \frac{\sqrt{\rm{A}_{\rm{d}}\Delta f}}{\rm{NEP}} 
\end{equation}
    \end{minipage}\vskip1em

where NEP is the incident optical power that results in a signal-to-noise ratio (SNR) of one, R is the responsivity of the device at a given wavelength and bias conditions, A$_{\rm{d}}$ is the photoactive device area and $\rm{\Delta}$f is the electrical bandwidth. The NEP is determined by the ratio of the root-mean-square (RMS) noise and responsivity R of the photodetector at a given wavelength. To extract NEP and D$^*$, the time-dependent response of the GSOI-planar device has been recorded with 8 Hz sampling rate at bias voltages of V$_b$ = 0 and 2V, respectively, under optical excitation at a wavelength of $\lambda$ = 520nm, chopped at 0.5s intervals. Fig.\ref{fig:noise}a) shows the IV characteristics of the GSOI-planar device in the dark and under periodic optical excitation with decreasing optical power at a bias voltage of V$_b$ = 0V. The RMS noise is extracted from the measurement in the dark, equating to 2.89pA. NEP and D$^*$ of the GSOI-planar device are calculated to be 14.5pW and 7.83$\times10^{\rm{10}}$ cm Hz$^{\rm{1/2}}$W$^{\rm{-1}}$, respectively. Under an applied reverse bias of V$_b$ = 2V, shown in Fig.\ref{fig:noise}b), the rise in the dark current leads to an increase in the RMS noise to 4.06pA. The NEP increases by $\sim$20\% to 17.7pW and D$^*$ decreases to 6.41$\times10^{\rm{10}}$ cm Hz$^{\rm{1/2}}$W$^{\rm{-1}}$. For comparison, our graphene-bulk silicon Schottky photodiodes exhibit a RMS noise of 40pA and NEP and D$^*$ of 190pW and 1$\times10^{\rm{10}}$ cm Hz$^{\rm{1/2}}$W$^{\rm{-1}}$, respectively. As such, a reduction of the photoactive silicon thickness from 500$\mu$m (bulk) to 10$\mu$m (SOI) leads to a $\sim$10 times improved NEP for the GSOI-planar device due to reduced noise. We anticipate that a passivation/encapsulation of the graphene surface will improve the sensitivity of GSOI photodetectors further. Passivation will prevent the adsorption of e.g. $\rm{H}_2$\rm{O} and $\rm{O}_2$ from the ambient environment onto graphene and reduce electronic noise \cite{Rumyantsev2010noiseg}. Additional sensitivity enhancement can be achieved through engineering of the graphene-silicon interface to further reduce dark currents\cite{li2016high,selvi2018towards}.

\begin{figure}[htbp]
\centering{
\includegraphics[width=85mm]{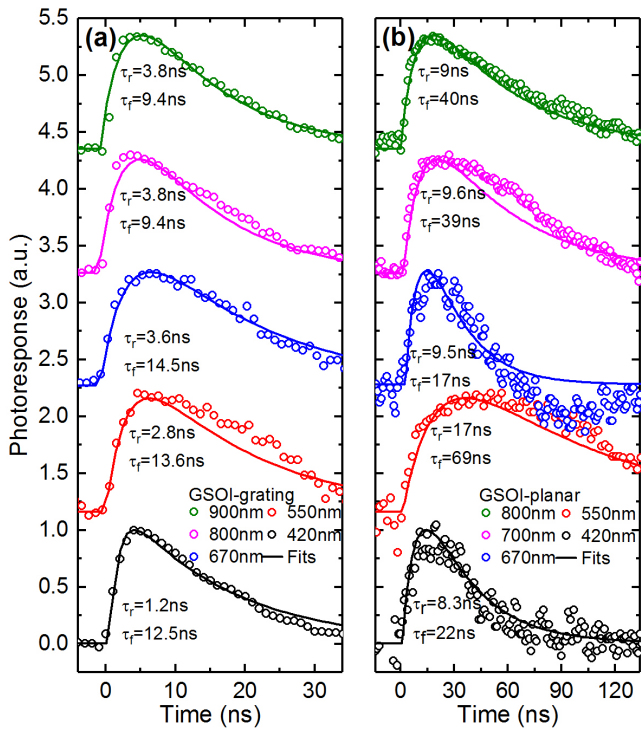}}
\caption{Time-resolved photoresponse of a) the GSOI-grating and b) the GSOI-planar device under illumination with a picosecond white light source at various wavelengths. Note change of scale of time-axis.}
\label{fig:speed}
\end{figure}

The high-speed optical response of both devices has been evaluated for wavelengths ranging from the near UV to NIR employing a picosecond white light laser source (Fianium, pulse duration $\tau_{\rm{pulse}} <$ 80ps at all wavelengths) and optical bandpass filters (FWHM $\leq$ 10nm). The electrical response was recorded using a transimpedance preamplifier (Phillips Scientific 6954) and an oscilloscope (WaveJet 354-A) while the devices were operated under zero bias conditions (V$_b$ = 0V). The current response I of the devices to the picosecond light pulses is shown in Figs.\ref{fig:speed})a,b) and has been fitted with a biexponential function

\begin{equation}
  I = I_{\rm{max}} \times (1-e^\frac{t-t_0}{\tau_r}) \times e^\frac{t-t_0}{\tau_f}
\end{equation}

with $I_{\rm{max}}$ the peak response, $\tau_r$ and $\tau_f$ the rise- and fall-times and $t_0$ the time-offset at which the photoresponse occurs. The rise- and fall-times of the GSOI-grating device are $\tau_r <$ 4ns and $\tau_f \sim$ 10-15ns for all tested wavelengths ($\lambda$ = 420-900nm) and are moreover almost constant for all wavelengths. The GSOI-planar device exhibits slightly increased rise- and fall-times in the orders of $\tau_r \sim$ 10ns and $\tau_f \sim$ 20-70ns, attributed to the larger average thickness of the active silicon layer of the GSOI-planar device compared to the GSOI-grating device with a patterned surface. Note that fall-times can be further decreased upon application of a bias voltage V$_b$ to facilitate extracting charge carriers out of the device; due to restrictions in our characterization setup this is not possible in present experiments.

The temporal response of a photodiode to an instant injection of excess carriers upon photon absorption depends mainly on three independent components which can be expressed\cite{goldberg1999semiconductor} by

\begin{equation}
\tau_{\rm{r}} = \sqrt{{\tau_{\rm{RC}}}^2 + {\tau_{\rm{dr}}}^2 + {\tau_{\rm{diff}}}^2}
\label{eq:responsetime}
\end{equation}

with $\tau_{\rm{r}}$ the overall 10\% to 90\% response time of the photodiode, $\tau_{\rm{RC}}$ the resistor-capacitor (RC) time constant of the photodiode, $\tau_{\rm{dr}}$ the electric-field driven drift time of the carriers in the depletion region and $\tau_{\rm{diff}}$ the diffusion time of carriers in the non-depleted region of the substrate. The wavelength dependent optical absorption coefficient of silicon determines the penetration depth of incident light into the silicon substrate according to Beer-Lambert law \cite{sze2007physics}. Generally, light of shorter wavelengths has a shallow penetration depth while light of longer wavelengths can penetrate much deeper into the bulk of the silicon substrate. In other words, incident short wavelength light optically excites charge carriers in the silicon close to the substrate surface while long wavelength creates charge carriers deep within the silicon substrate. Considering a depletion region length $L_D$ of the Schottky junction for employed low doped silicon on the order of 1-2 $\mu$m \cite{selvi2018towards}, this implies complete absorption of short wavelength light (e.g. $\lambda$ = 450nm) within the depletion region while only $\sim$50\% and $\sim$15\% of light of $\lambda$ = 550nm and $\lambda$ = 700nm wavelengths, respectively, are absorbed within the depletion region. The light absorbed deeper within the silicon substrate, outside the depletion region length of the Schottky junction, thus creates charge carriers that contribute to the photoresponse by a diffusion process. The transit time for charge carriers within the depletion region for zero external bias condition depends on the depletion region length $L_D$ and the electric field E due to the built in potential V$_{\rm{bi}}$ of the Schottky junction and can be estimated by $\tau_{\rm{dr}}= \frac{L_D} {\mu E}= \frac{L^{2}} {\mu V_{\rm{bi}}}$ to $\sim$82ps for holes and $\sim$26ps for electrons based on mobilities for holes and electrons of $\mu_p$ = 450$\frac{cm^2}{Vs}$ and $\mu_n$ = 1400$\frac{cm^2}{Vs}$, respectively.
However, charge carriers contributing to diffusion currents from deep within the silicon substrate need to travel far longer distances. The required time for diffusion can be calculated via  $\tau_{\rm{diff}}= \frac{L_{\rm{p}}^{2}} {2D_{\rm{p}}}$. In the case of a bulk silicon substrate, the diffusion length $L_{\rm{p}}$ for holes is determined by the thickness of the active silicon layer. The diffusion coefficient $D_{\rm{p}} = \mu_{\rm{p}} \frac{kT}{q}$ can be evaluated from the mobility of holes ($\mu_{\rm{p}} \sim 450 \frac{cm^2}{Vs}$) in lowly n-doped silicon. Using a recombination time of $\tau_{\rm{p}} = 2\times10^{-4}s$ for n-type silicon with a doping level of N$_d$ = 3.5$\times10^{14}$ cm$^{-3}$ , the diffusion length $L_{\rm{p}}$ is calculated as $L_{\rm{p}} = \sqrt{D_{\rm{p}} \tau_{\rm{p}}} = 480 \ \mu m$ which is comparable to the thickness of a standard silicon substrate.

In a straight forward manner, the necessary time $\tau_{\rm{diff}}$ for a hole to diffuse 10$\mu$m and 480$\mu$m, representative of employed SOI wafers and typically employed bulk silicon wafers with identical doping concentration of N$_d$ $\sim$3.5$\times10^{14}$ cm$^{-3}$, can be calculated as $\sim$42ns and $\sim$98$\mu$s, respectively. As such we argue that the decrease of the active optical silicon layer thickness to 10$\mu$m in our devices based on SOI material and with it a reduction of speed limiting diffusion currents is the major reason for an increase of the operating speed of both the GSOI-planar and GSOI-grating device by more than two orders of magnitude compared to GS devices based on bulk silicon.

\begin{figure}[htbp]
\centering{
\includegraphics[width=80mm]{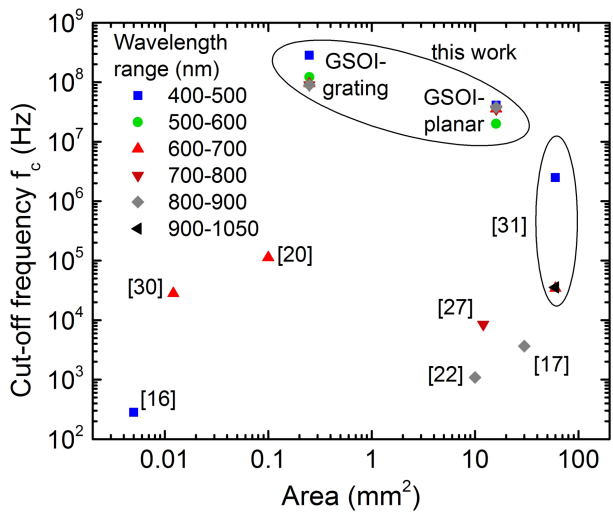}}
\caption{Comparison of the 3dB cut-off frequency f$_c$ versus photoactive device area of our GSOI-photodetectors with GS Schottky photodetectors fabricated on bulk silicon photodiodes at various wavelengths.}
\label{fig:cut-off}
\end{figure}

The argument of diffusion currents being the speed limiting component is further supported by comparing the 3-dB cut-off frequency (f$_{\rm{c}}$) of our GSOI devices fabricated on SOI substrate with GS junction diodes fabricated on bulk silicon substrates in the literature, including our previous report (Fig.\ref{fig:cut-off}) \cite{selvi2018towards}. The corresponding cut-off frequencies of the GSOI-grating, GSOI-planar and reference devices are determined from the measured and reported rise times ${\tau}_{\rm{r}}$ as f$_{\rm{c}}$ = 0.34/$\tau_{\rm{r}}$ \cite{liuphotonic}. Fig.\ref{fig:cut-off} shows the cut-off frequency (f$_{\rm{c}}$) vs photoactive device area for various wavelengths of our own devices and devices reported in the literature. It can be seen that despite an active area of our devices comparable to or greater than devices fabricated on bulk silicon, our devices fabricated on SOI substrate exhibit an increase in cut-off frequency of several orders of magnitude. It further demonstrates that the increase in cut-off frequency f$_{\rm{c}}$ of our devices cannot be attributed to a simple decrease of the RC-time constant due to down scaling contact and junction areas. To the best of our knowledge, our graphene-silicon Schottky photodetectors on SOI substrate are the fastest to date for free space light detection, especially in the VIS and NIR wavelength range $\lambda >$ 500nm.

\begin{figure}[htbp]
\centering{
\includegraphics[width=80mm]{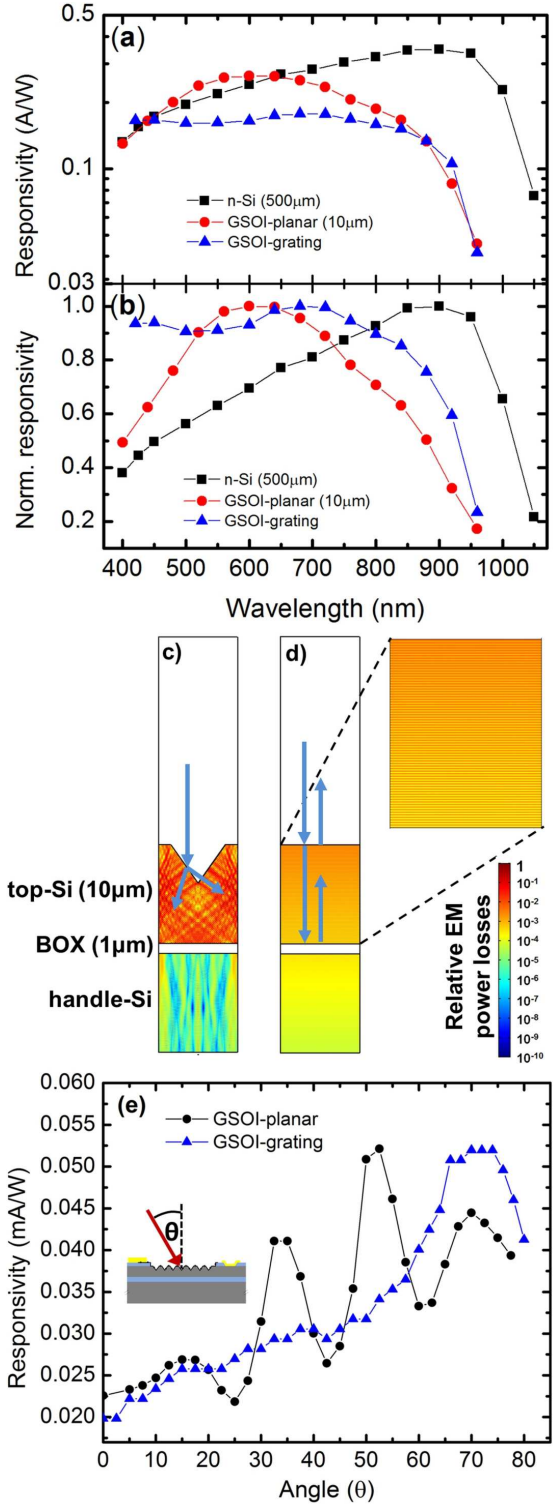}}
\caption{Effect of substrate thickness, surface topography, and incident angle on the spectral response of GS Schottky photodiodes. a) Spectral responsivity of the GSOI-planar and -grating device and comparison with a GS Schottky diode fabricated on a $\sim$ 500$\mu$m thick silicon substrate. b) Comparison of normalized responsivity of GSOI-planar, GSOI-grating and bulk GS diode. c,d) Simulated relative electromagnetic power loss density in the grating and planar device, respectively, for normal incidence light and at a wavelength of $\lambda $ = 980nm. Blue arrows indicate the light reflection and transmission, respectively. Zoom-in: Interference effects in the planar device. e) Angular dependence of the responsivity of the GSOI-planar and -grating device, respectively.}
\label{fig:Spectral}
\end{figure}

The spectral response of the photodetectors on SOI substrate in comparison with a device fabricated on 500$\mu$m thick bulk n-type silicon with identical doping level is shown in Fig.\ref{fig:Spectral}a). All three device types show similar responsivities for shorter wavelengths $\lambda <$ 450nm since incident light in this wavelength range is totally absorbed within the first top 10$\mu$m of silicon. Light of longer wavelengths $\lambda >$ 450nm can penetrate deeper into the silicon substrate due to the reduced absorption coefficient. For longer wavelengths, only part of the incident light is absorbed within the active top silicon layer of the SOI substrate and the remaining light power is transmitted into the BOX and silicon handle wafer where it does not contribute to the photoresponse. This is manifested in Fig.\ref{fig:Spectral}a) as a reduced responsivity of the GSOI devices compared to GS devices fabricated on bulk silicon for longer wavelength light $\lambda >$ 900nm. It is further noticeable that the spectral response of the GSOI-grating device is almost flat in the wavelength range $\sim$ 400-800nm compared to both the GSOI-planar and bulk silicon device, easier recognizable in the plot of the spectral dependence of the responsivities for each device type normalized to their respective peak responsivities (Fig.\ref{fig:Spectral}b)). This flat spectral photoresponse of the GS-grating device can be attribute to the surface topography of the device. The V-grooves form micro-optical elements that allow direct transmission of incident light into the silicon substrate but also reflect light that is unused in planar devices back into the substrate \cite{Chong2012trapping,Wang2012trapping,Haase2007trapping}. Fig.\ref{fig:Spectral}c,d) show the simulated electromagnetic power loss for a unit cell of the grating and planar device, respectively, for normal incident light (p-polarized) at a wavelength of $\lambda$ = 980nm (ESI). The blue arrows are indicate the transmitted and reflected light paths. Light incident on the GSOI-grating device with 54.7$^\circ$ sloped facets will be partially transmitted into the silicon substrate after undergoing refraction. However, the reflected light component is directed towards the opposite facet and also partially absorbed by the substrate. Additionally, the active silicon layer of the SOI substrate of both the GSOI-grating and the GSOI-planar device forms an optical cavity where multiple reflections between the topmost air-silicon, silicon-BOX and BOX-handle silicon interfaces lead to optical interference effects. This is clearly visible in the zoom-in on the electromagnetic losses in the GSOI-planar device (Fig.\ref{fig:Spectral}d)). The surface topography of the GSOI-grating device leads to greater electromagnetic power losses (absorption) within the active silicon layer compared to the GSOI-planar device and allows tailoring the spectral response of GS photodetectors. However, we note that optical absorption on its own does not fully describe the electrical responsivity. Optically excited charge carriers contribute to the photocurrent via diffusion and drift processes subject to electric fields in the formed junction which can be particularly complex, especially for the grating device.

Devices fabricated on SOI substrate are further sensitive to the angle of incident light. Fig.\ref{fig:Spectral}e) shows a measurement of the angular dependence of the responsivity for the GSOI-planar and GSOI-grating device, respectively, for p-polarized light of $\lambda$ =  980nm wavelength. The GSOI-planar device, in which all interfaces between different materials are parallel, exhibits a strong angular dependence. Constructive and destructive interference effects due to multiple reflections within the optical cavities formed by the active silicon layer and the BOX result in an oscillating behavior of the responsivity. The GSOI-grating device, however, does not show this oscillatory behavior but instead exhibits a smoother angular dependence and an increased responsivity at higher incident angles due the surface topography of the device. The micro-optical elements formed by the slanted facets allow more efficient coupling of incident light into the silicon substrate at higher incident angles. Furthermore, parallel air-top silicon and top silicon-BOX interfaces that would result in well defined interference effects are strongly reduced in the GSOI-grating device. Light entering the top silicon through the slanted surfaces can be incident on the top silicon-BOX interface above the critical angle of $\sim$ 25$^\circ$ which results in total internal reflection (TIR) of the light rays at this interface.

In conclusion, we have demonstrated that graphene-silicon Schottky photodetectors fabricated on an SOI substrate exhibit a significant improvement in operating speeds compared to their bulk silicon counterparts, approaching cut-off frequencies of f$_c$ $\sim$ 1GHz. The speed improvement can be attributed to a suppression of speed limiting diffusion currents due to the reduced thickness of the photoactive silicon layer. We anticipate that down scaling of the junction area to decrease the RC-time constant can further improve operating speeds. Ultimately, the thickness of the silicon substrate determines the trade off between responsivity and speed for GSOI-PDs. We demonstrated that an active silicon layer thickness of 10$\mu$m significantly improves the speed of the GSOI device with only a modest decrease in the responsivity. The integration of micro-optical elements through surface patterning allows control of the spectral responsivity and angular dependence of GSOI devices. These micro-optical elements might further pave the way towards GSOI photodetectors operating in a total internal reflection (TIR) architecture, particularly for wavelengths beyond $\lambda >$ 1.1$\mu$m for which silicon is fully optical transparent and absorption is taking place in the graphene only.

\section{Acknowledgements}

H.S. acknowledges funding from the Turkish government (MEB-YLSY). P.P. acknowledges funding from the Royal Society (RG140411). T.J.E acknowledges funding from the Huawei Innovation Research Program (HIRP).

\end{document}